\documentclass{template}
\usepackage{xcolor}
\usepackage{amsmath}
\usepackage[utf8]{inputenc}  
\DeclareUnicodeCharacter{00ED}{\'i}   
\DeclareUnicodeCharacter{00F1}{\~n}  
\DeclareUnicodeCharacter{00F3}{\'o}  
\DeclareUnicodeCharacter{0301}{\'{}} 
\DeclareUnicodeCharacter{00E1}{\'a} 
\DeclareUnicodeCharacter{00E9}{\'e} 

\begin{document}

\pagestyle{fancy}

\title{Integration of Cobalt Ferromagnetic Control Gates for Electrical and Magnetic Manipulation of Semiconductor Quantum Dots}

\maketitle


\author{Fabio Bersano*}
\author{Michele Aldeghi}
\author{Niccolò Martinolli}
\author{Victor Boureau}
\author{Thibault Aboud}
\author{Michele Ghini}
\author{Pasquale Scarlino}
\author{Gian Salis}
\author{Adrian M. Ionescu}



\begin{affiliations}
F. Bersano, N. Martinolli, T. Aboud, Dr. M. Ghini, Prof. A. M. Ionescu\\
Nanoelectronic Devices Laboratory (Nanolab), IEM EPFL, 1015 Lausanne, Switzerland \\ 
Email Address: fabio.bersano@epfl.ch, adrian.ionescu@epfl.ch\\
M. Aldeghi, Dr. G. Salis \\
IBM Research-Zurich, 8803 R{\"u}schlikon, Switzerland \\
Dr. Victor Boureau\\
Interdisciplinary Center for Electron Microscopy, EPFL, 1015 Lausanne, Switzerland\\
Prof. P. Scarlino \\
Hybrid Quantum Circuits Laboratory (HQC Lab), IPHYS EPFL, 1015 Lausanne, Switzerland 
\end{affiliations}



\justifying
\begin{abstract}
The rise of electron spin qubit architectures for quantum computing processors has led to a strong interest in designing and integrating ferromagnets to induce stray magnetic fields for electron dipole spin resonance (EDSR). The integration of nanomagnets imposes however strict layout and processing constraints, challenging the arrangement of different gating layers and the control of neighboring qubit frequencies. This work reports a successful integration of nano-sized cobalt control gates into a multi-gate fully-depleted silicon-on-insulator (FDSOI) nanowire with nanometer-scale dot-to-magnet pitch, simultaneously exploiting electrical and ferromagnetic properties of the gate stack at nanoscale. The electrical characterization of the  multi-gate nanowire exhibits full field effect functionality of all ferromagnetic gates from room temperature to 10 mK, proving quantum dot formation when ferromagnets are operated as barrier gates. The front-end-of-line (FEOL) compatible integration of cobalt is examined by energy dispersive X-ray spectroscopy and high/low frequency capacitance characterization, confirming the quality of interfaces and control over material diffusion. Insights into the magnetic properties of thin films and patterned control-gates are provided by vibrating sample magnetometry and electron holography measurements. Micromagnetic simulations anticipate that this structure fulfills the requirements for EDSR driving for magnetic fields higher than 1 T, where a homogeneous magnetization along the hard magnetic axis of the Co gates is expected. The FDSOI architecture showcased in this study provides a scalable alternative to micromagnets deposited in the back-end-of-line (BEOL) and middle-of-line (MOL) processes, while bringing technological insights for the FEOL-compatible integration of Co nanostructures in spin qubit devices.

\end{abstract}


\section{Introduction}
Throughout the development of von Neumann processing units based on nano-CMOS technology, the integration of thin ferromagnetic films like Co and CoFeB has a rich historical background. The applications spans from hard-disk drives to advanced spintronic devices based on spin-dependent tunneling effects, such as spin-transfer and spin-orbit torque magnetic random access memories (STT-MRAM, SOT-MRAM) \cite{finocchio2023roadmap}. More recently, the non-volatility of thin films magnetization has been explored for unconventional computing hardware, such as neuromorphic, in-memory, and quantum computing \cite{leonard2022shape,jung2022crossbar, watson2018programmable}. 
In the field of quantum computing with spins, academic research has introduced ferromagnets into quantum dot architectures to enable electron spin manipulation \cite{pioro2008electrically, yoneda2018quantum, kawakami2014electrical, leon2020coherent, zhang2021controlling, philips2022universal, yoneda2015robust, bersano2023quantum}. Simultaneously, significant efforts are being made to develop modular CMOS-compatible manufacturing processes for spin qubits at the industrial level \cite{sabbagh2019quantum, veldhorst2017silicon, gonzalez2021scaling, zwerver2022qubits}. These advancements underscore the importance of investigating the integration of ferromagnets in back-end-of-line (BEOL) \cite{klemt2023electrical,koch2024industrial300mmwaferprocessed} and front-end-of-line (FEOL) compatible processes \cite{aldeghi2023modular}.

In the case of Loss-DiVincenzo electron spin qubits \cite{loss1998quantum}, the integration of ferromagnetic elements was originally proposed to serve as spin-polarized fermionic baths for the initialization of spin states. In real practice, spin selective tunneling protocols between electron reservoirs and quantum dots \cite{elzerman2004single, philips2022universal} are most commonly employed. Nevertheless, ferromagnets play an important role in implementing single electron spin qubit gates. Coherent electron spin manipulation can be controlled using oscillatory electric fields in an inhomogeneous magnetic field \cite{rolf_gian_patent}, an approach known as electric dipole spin resonance (EDSR), where a synthetic spin-orbit coupling (SOC) is induced by magnetic field gradients shaped by purposely engineered micromagnets. This technique has two main advantages compared to electron spin resonance (ESR): it requires only AC electric fields (instead of AC magnetic fields) which are considerably easier to engineer, and eliminates the Joule heating induced by microstrip lines, enabling faster spin control (i.e., higher Rabi frequencies) \cite{veldhorst2014addressable, wang2022ultrafast}. However, due to the typical micro/nanometer size of ferromagnets, EDSR driven by magnetic gradients has been considered unfavorable for large-scale integration of electron spin qubits \cite{camenzind2022hole, gilbert2023demand, liu2024progress}. A second concern comes from simple magnetostatic considerations: as a specific driving gradient is maximized to enable high Rabi frequencies, high decoherence gradients are automatically induced, worsening qubit coherence. Therefore, precise control of the magnetic field profile at the nanoscale and the miniaturization of magnets are essential for developing high-performance, scalable spin qubits \cite{liu2024progress}.

Particularly for electron spin qubits in Si, the need for plunger, depletion/barrier gates and micromagnet integration puts stringent restrictions on layout level when scaling up the quantum processor, increasing the complexity of gating layers for electrostatic control. This challenge can be addressed by properly designing the device architecture and combining the magnetic and electrical functionality of ferromagnetic control gates \cite{forster2015electric,tadokoro2021designs,bordoloi2022spin}. While a 2D architecture proves more fitting for expansive networks \cite{borsoi2024shared}, it necessitates additional depletion gates to induce 1D confinement, a role that is naturally fulfilled by etched structures like FinFETs and nanowires, which typically require a minimum of three metal layers \cite{camenzind2022hole, philips2022universal}. 

\begin{figure}
\centering
  \includegraphics[width=0.8\linewidth]{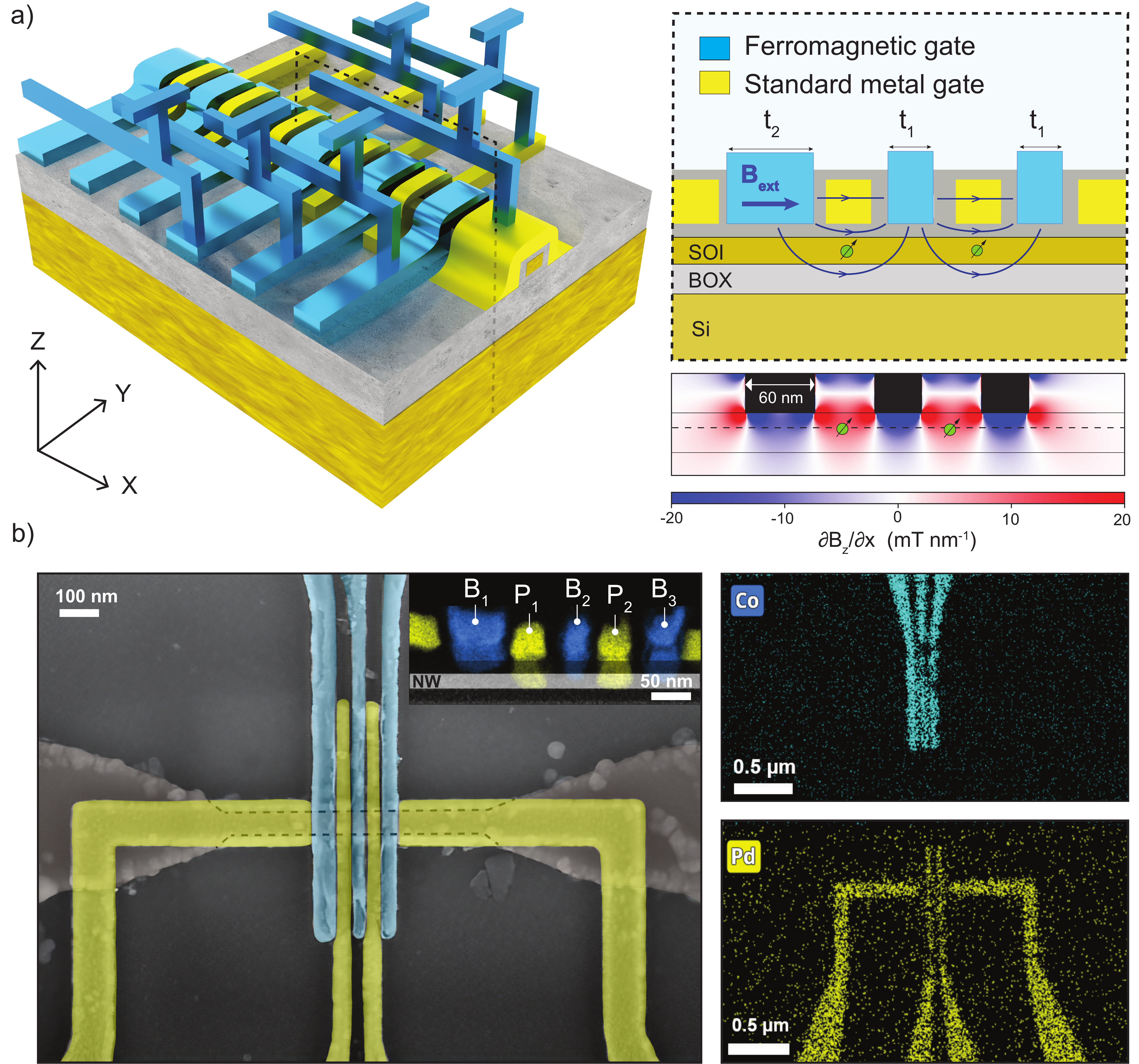}
  \caption{a) Schematic of an FDSOI nanowire architecture featuring ferromagnetic gates for electrical and magnetic control of quantum dots with integrated MOL cobalt interconnects, and micromagnetic simulation results for the magnetic field gradient generated along the nanowire cross-section by the Co gates (in black) for an external magnetic field of 1 T oriented along the $x$-direction (the estimated position of the quantum dots is highlighted in green). b) SEM image of a fabricated double-dot device featuring cobalt barrier gates (B$_1$, B$_2$, and B$_3$) and palladium plunger gates (P$_1$ and P$_2$), with EDX-SEM analysis of the materials. }
  \label{fig:figure_1}
\end{figure}

In this work, we propose a device architecture designed to address the scalability challenges of EDSR-driven electron spin qubits, enhancing both driving gradient and addressability of neighbouring qubits, while minimizing decoherence gradients. In our structure, the 1D confinement of the quantum dots is achieved using ultra-thin fully-depleted silicon-on-insulator (FDSOI) nanowires. Ferromagnetic gates are integrated to produce slanting Zeeman fields while controlling the local depletion of minority carriers within the SOI nanowires through a DC bias, which could be combined with RF excitation (\textbf{Figure \ref{fig:figure_1}a}).
Two metal bi-layers, Ti/Pd and Cr/Co, act as plunger and barrier gates, respectively. Palladium is chosen for its small grain size \cite{lawrie2020quantum}, which facilitates patterning via lift-off, and for its resistance to self-oxidation, unlike aluminum. Cobalt is selected as the ferromagnetic element due to its compatibility with industrial processes for integrated circuits and higher saturation magnetization compared to nickel.
Although Fe and its permalloys exhibit higher magnetic saturation and typically lower magnetocrystalline anisotropy than Co \cite{cullity2011introduction, hubert2008magnetic}, they are excluded due to iron being considered one of the most problematic contaminants in metal-oxide-semiconductor devices \cite{istratov2000iron}. The incorporation of Co nano-gates suggested here is first analyzed with an emphasis on its compatibility with FEOL and middle-of-line (MOL) CMOS processes. The fabrication and electrical characterization of a proof-of-concept quantum dot device (\textbf{Figure \ref{fig:figure_1}b}) is reported, and micromagnetic simulations are discussed to anticipate the experimental conditions required for EDSR driving induced by Co barrier gates.

\section{FEOL-compatible Cobalt Integration}
\subsection{Thin Film Morphology and Thermal Budget}
In the context of semiconductor devices manufacturing, cobalt is considered a CMOS-compatible material. It has been explored as a replacement for traditional interconnects metals like tungsten and copper due to its low electron mean free path \cite{narasimha20177nm}, and as a valid alternative to TiSi$_2$ and NiSi to form low-resistivity silicide contacts (CoSi$_2$) in advanced technology nodes \cite{ichimori2002fully}. Co interconnects have been successfully integrated for CMOS nodes below 100 nm in the MOL metallization layers and as cladding coating at higher metal levels, showing more than two times reduction in via resistance \cite{kamineni2016tungsten} and 5-10 times improvement in electromigration \cite{auth201710nm}. However, instances of extrinsic time-dependent dielectric breakdown and atomic diffusion have been documented, prompting the need for diffusion barriers \cite{el2010inter, dash2008near}. Even a partial diffusion of Co atoms into the gate oxide and silicon could result in the creation of interfacial silicides \cite{peng2006adsorption}, which may lead to the development of local magnetic moments with random orientations compared to the magnetization of the thin film. Yet, it has been demonstrated that a thin SiO$_2$ layer effectively inhibits the formation of interfacial CoSi/CoSi$_2$ due to the strong binding of Si atoms to the thermal SiO$_2$ layer \cite{vcechal2008morphology}. However, aluminum oxide does not exhibit this diffusion barrier effect \cite{el2010inter}, necessitating the use of an additional adhesion layer to separate the Co from the gate oxide. Because of the partial diffusion of Cr into high-\textit{k} dielectrics and reaction with oxygen to form Cr$_2$O$_3$, a Cr-Co alloy provides a reliable self-forming diffusion barrier that shows a breakdown voltage two times higher than pure Co \cite{kim2022robust}. Motivated by this, we first deposit a thin Cr layer (2-3 nm) before Co e-beam evaporation; this layer also enhances metal adhesion to the gate oxide and improves the lift-off process. The cross-section of a Cr/Co barrier gate and Ti/Pd plunger gate annealed for 10 minutes at 300 °C in forming gas (5$\%$ H$_2$, 95$\%$ N$_2$) are shown in \textbf{Figure 2a}, where energy dispersive X-ray (EDX) spectroscopy in a transmission electron microscope (TEM) was used to highlight the materials. No Cr/Co or Ti/Pd diffusion was observed in the oxide, indicating effective atomic screening by the thin adhesion metal, though partial diffusion of the adhesion layers into Co and Pd occurred. 

\begin{figure}
\centering
  \includegraphics[width=\linewidth]{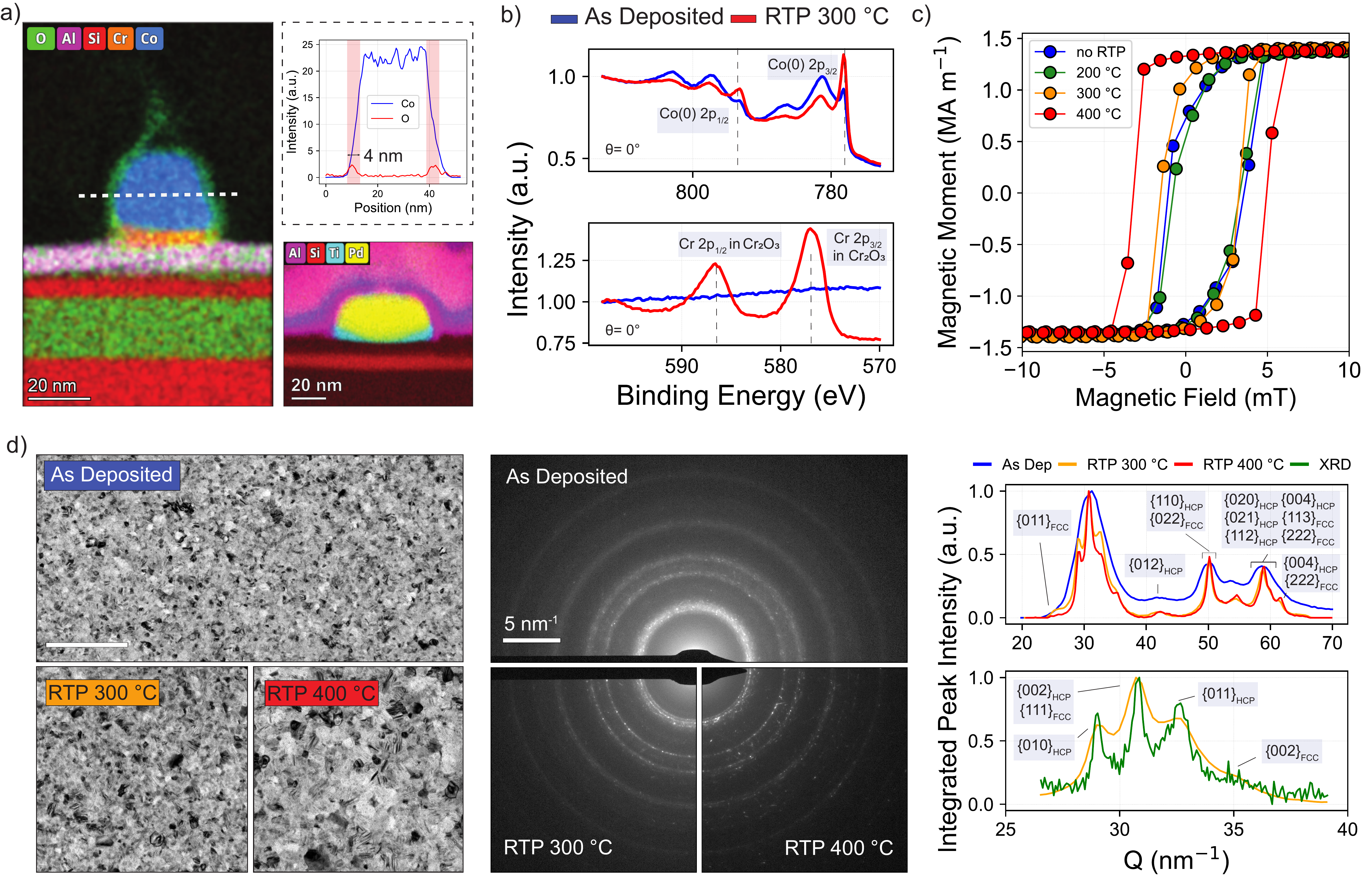}
  \caption{Material analysis. a) EDX-TEM analysis of Cr-Co and Ti-Pd gates on FDSOI with a cutline highlighting the surface oxidation of cobalt (sample annealed for 10 minutes in forming gas at 300\;°C). b) XPS results for e-beam evaporated Cr/Co thin films as-deposited and annealed at 300\;°C in forming gas. $\theta$ is relative to the surface normal. c) VSM measurements of the same films annealed at different temperatures. d) Bright-field TEM micrographs and SAED patterns for the as-deposited and annealed Co thin films, with azimuthal integration of the Bragg peak intensities vs scattering vector to highlight the crystal phase and overlap with GIXRD measurement. The scale for the bright-field images is $200$ nm.
  }
  \label{fig:figure_2}
\end{figure}

In addition to inter-layer diffusion, the oxidation of Co must be considered since it occurs rapidly when in contact with oxygen \cite{smardz1992oxidation}. Three oxide growth processes have been documented: an immediate hydroxide formation (Co(OH)$_2$) of a few \r{A}ngstrom ($< 1$ nm) upon air exposure of the surface at annealing temperatures below 100\;°C, a logarithmic growth of CoO for annealing temperatures between 100-225 °C, and a quadratic growth rate of a combination of CoO and Co$_3$O$_4$ for temperatures exceeding 300 °C \cite{tompkins1981oxidation, fontaina2014room}. We analyzed the oxidation phase of Co through angle-resolved X-ray photoelectron spectroscopy (ARXPS) of continuous thin films (3/33 nm of Cr/Co) deposited on both SiO$_2$ and Al$_2$O$_3$ before and after rapid thermal annealing (RTP). Results are independent of the dielectric used and are reported in \textbf{Figure \ref{fig:figure_2}b} for $\theta=0\degree$ (detection normal to the surface). Upon collection of a high resolution spectrum in the interval typical of cobalt, two sets of peaks separated by $\Delta \approx 15$ eV are measured, identifying the binding energy profile of Co and its oxides in accordance with data reported in literature \cite{chastain1992handbook}, with a sharp increase in the metallic peak measured after RTP as typical of surface dehydration. The detection at lower binding energies identifies two peaks corresponding to Cr$_2$O$_3$ in the sample subjected to RTP at $300$ °C, providing experimental evidence of Cr diffusion into Co and its segregation at the outer interface. This is further confirmed by a quantitative analysis at higher detection angles and EDX-TEM for a nanogate cross-section (details in SI). Vibrating sample magnetometry (VSM) measurements were conducted to study the evolution of coercivity with rapid thermal treatments in forming gas, results are shown in \textbf{Figure \ref{fig:figure_2}c}. In agreement with previous studies on thermally induced changes of structural and magnetic properties in Co thin films \cite{jergel2009annealing}, an abrupt increase in coercive field is measured at an annealing temperature of 400\;°C. This behavior was previously attributed to improvements in the crystalline structure and magneto-crystalline anisotropy of the growing \textit{fcc} crystallites, along with enhanced roughness at the metal-dielectric interface. We studied the evolution of crystallite size and crystal phase of thin Co evaporated films by bright-field (BF) TEM and selected area electron diffraction (SAED) measurements \cite{klinger2015crystallographic,klinger2017more}. The results, shown in \textbf{Figure \ref{fig:figure_2}d}, indicate a grain size ranging from 15 to 20 nm, which remains mostly unaffected by the thermal treatment at $300$ °C, but roughly doubles in size at $400$ °C. The higher intensity peaks correspond to the hexagonal close packed (\textit{hcp}) phase, as confirmed by grazing incidence XRD (GIXRD) measurements. After annealing at 300\;°C, an improvement in the crystalline structure and stress relaxation are attained, as suggested by the reduction in diffuse brightness around the main peaks, which in turn highlights the presence of face centered cubic (\textit{fcc}) phase. The \textit{fcc} phase appears to slightly increase in the sample annealed at $400$ °C, and this, together with the crystallites growth and possibly changes in the interface, could contribute to explain the observed rise in magnetic coercivity. The decreased uniformity in the SAED rings has to be attributed to the smaller number of crystallites framed by the same BF field of view, rather than to any film texturing.

Both metallization layers of the multi-gate architecture proposed in this work are patterned on thin dielectrics deposited via atomic layer deposition (ALD) at $300$ °C. We observed a high risk of breaking the nanogates when annealed at temperatures above the dielectric deposition temperature, due to stress generated at the dielectric-metal interface. During all annealing processes conducted up to 300 °C, surface oxidation of the Co nanogates was limited to 4 nm by cooling the sample to room temperature in a nitrogen atmosphere before exposing it to ambient conditions. Our study thus recommends 300 °C as a safe thermal limit for integrating Co nanomagnets on ALD-dielectrics (which are typically deposited at lower temperatures), as it minimizes the risk of magnetic inhomogeneities caused by high oxidation rates and interfacial stress between the ferromagnets and dielectric. Further characterization studies should explore the feasibility to extend the thermal budget to $400–450$ °C (typical for CMOS back-end processing) by patterning Co nanogates on thermally oxidized SOI nanowires.

\subsection{Work Function and Interface Traps}
\begin{figure}
\centering
  \includegraphics[width=\linewidth]{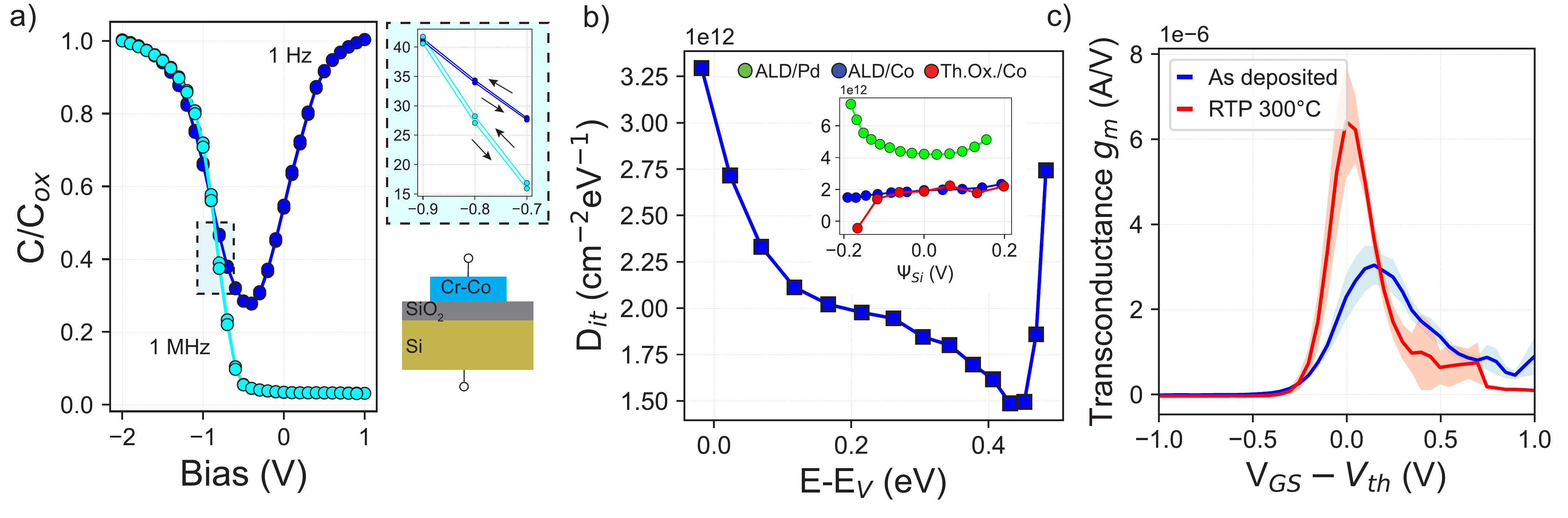}
  \caption{Characterization of Co-FETs. a) High and low frequency C-V measurements of a Cr-Co/SiO$_2$ MOS-capacitor, with a zoom-in to highlight the minimal hysteresis induced by the gate stack. b) Extracted density of interface traps within the silicon band-gap for a Cr-Co/ALD-Al$_2$O$_3$ MOS-capacitor (the same gate stack of the fabricated multi-gate FDSOI nanowire). The inset showcases a comparison of $D_{it}$ between different gate/oxide layers around the Si flat-band potential. c) Transconductance value extracted for 60 Co-FETs measured before and after thermal treatment, the shaded colors indicate the standard deviation around the mean value.}
  \label{fig:figure_3}
\end{figure}

The primary criteria for incorporating a metal as gating layer include its chemical stability with a given dielectric to regulate the threshold voltage (V$_{th}$), negligible C-V hysteresis, and low interface traps. All these properties depend on the thermal history of the fabrication process and set boundaries to the thermal budget. We analyzed the density of traps in metal-oxide-semiconductor (MOS) capacitors fabricated with Cr-Co (3-33 nm) and Ti-Pd (3-33 nm) on thermal SiO$_2$ (grown at 850 °C in dichloroethene atmosphere) and Al$_2$O$_3$ deposited by ALD (with trimethylalluminum and H$_2$O precursors) using frequency dependent capacitance measurements, where the back side of the samples was coated with a silver paste to form a better contact. The Ti-Pd bilayer has been extensively investigated for gating applications in the context of sensitive electronic devices (e.g., electron and hole spin qubits \cite{boter2020scaling, brauns2018palladium}), and facilitates evaporation and lift-off processes while exhibiting low charge noise. Consequently, it serves as a valuable benchmark for a comparison with the Cr-Co bilayer. An example of C(f)-V measurements for a cobalt MOS with thermal SiO$_2$ is reported in \textbf{Figure \ref{fig:figure_3}a}. The effective metal work function ($\phi_{m,eff}$) and density of traps (N$_{eff}$) have been computed by comparing the high frequency characteristics at 1 MHz for different oxide thicknesses ($t_{ox}$) using the $1/C_{tot}^2$ method \cite{nicollian2002mos} and Equation (\ref{eq:1}), where the flat-band voltage (V$_{FB}$) is extracted from the experimental total capacitance C$_{tot}$(V$_G)$ measured as a function of the applied bias (V$_G$). C$_{ox}$ is the oxide capacitance measured in accumulation, $N_a$ the doping of the substrate, $q$ the elementary charge, $\phi_s$ the semiconductor potential, and $\epsilon_{Si/ox}$ the dielectric constant of silicon/oxide, respectively:
\begin{align}
    \begin{cases}
        \frac{1}{C_{tot}(V_G)^2 }= \frac{1}{C_{ox}^2} + \frac{2}{q N_a \epsilon_{Si}}\left(V_G - V_{FB}\right) \\
        V_{FB} = \phi_{m,eff} - \phi_s + \frac{N_{eff} t_{ox}}{\epsilon_{ox}}
    \end{cases}
    \label{eq:1}
\end{align}
Each measurement was repeated on at least 4 devices with and without thermal annealing; average values are summarized in \textbf{Table 1}. 
There are two main observations: first, the effective work function of metals shows a clear dependence on the dielectric and annealing conditions; second, the density of charge traps is of the same order of magnitude ($10^{11}$ cm$^{-3}$) for Ti/Pd and Cr/Co gates, and is reduced in both cases after thermal treatment. The value of $\phi_{m,eff}$ is closer to the vacuum condition for SiO$_2$ whereas it is higher for ALD-Al$_2$O$_3$ gate stacks. This observation is supported by the interface dipole theory, which suggests that electric dipoles form at the interface due to nonstoichiometric interfacial regions (introducing charge traps) and band offsets at the metal-dielectric interface \cite{gu2006effective}. The effective metal work function determined from C-V measurements of a MOS heterostructure differs from its value in vacuum ($\phi_m$) by the following relationship:
\begin{equation}
    \phi_{m,eff} = E_{CNL,d} + S(\phi_{m}-E_{CNL,d}),    
   \label{eq:2} 
\end{equation}
where $S$ is the screening parameter of the dielectric, determined from the electronic component of the dielectric constant ($S = [1+0.1(\epsilon_{\infty}-1)^2]^{-1}$) \cite{yeo2002effects}, and $E_{CNL,d}$ is the charge neutrality level of the dielectric. Both $S$ and $E_{CNL,d}$ depend on the fabrication process and are responsible for the pinning of $\phi_{m,eff}$ due to band alignment at the metal-dielectric interface. An higher pinning effect is expected for high-$k$ dielectrics compared to SiO$_2$ due to a smaller slope $S$. In the case of ALD-Al$_2$O$_3$ deposited on Si, $E_{CNL,Al_2O_3}$ is generally below the mid-gap position \cite{zhou2022experimental, monch2007electric}, and therefore expected to be lower than the Fermi level of Ti/Pd and Cr/Co bilayers with respect to the level of vacuum. This indicates that negative charge transfer occurs at the interface, driving the band lineup and finally inducing electric dipoles that lead to an higher effective measured work function \cite{gu2006effective}. The increase in work function is even more pronounced after annealing in forming gas, which enhances interlayer metal diffusion \cite{lu2010bilayer} and is expected to lower $E_{CNL-Al_2O_3}$ \cite{zhou2022experimental}. 

The density of interface traps ($D_{it}$) in field-effect devices depends both on the oxide and the top gating layers. To attain a deeper understanding of the impact of Cr-Co double-layer on interface charges, the density of interface traps in the band-gap of the $p$-type silicon substrate has been extracted for both Ti/Pd and Cr/Co using the high-low frequency comparison method \cite{nicollian2002mos}. $D_{it}$ is computed from Equation (\ref{eq:3}): 

\begin{equation}
    D_{it}(\psi_s) = \frac{1}{qA}\left[\left(\frac{1}{C_{lf}(\psi_s)}-\frac{1}{C_{ox}}\right)^{-1}-\left(\frac{1}{C_{hf}(\psi_s)}-\frac{1}{C_{ox}}\right)^{-1}\right],
    \label{eq:3}
\end{equation}
where $C_{lf}(\psi_s)$ and $C_{hf}(\psi_s)$ are the low and high frequency measured total capacitances, $q$ the elementary charge, $A$ the area of the capacitor, and $C_{ox}$ the oxide capacitance measured as the total capacitance in accumulation. The semiconductor potential dependence on the gate voltage $V_G$ is computed by numerical integration of the measured capacitance:

\begin{equation}
    \psi_s(V_G)- \psi_s(V_0) = \int_{V_0}^{V_G} \left[ 1- \frac{C_{lf}(V)}{C_{ox}}\right] dV,
\end{equation}
and can be referred to the distance from the silicon valence band edge according to:

\begin{equation}
    E(V_G)-E_V = \frac{E_G}{2}-q\psi_s(V_G)- \frac{kT}{q}ln\left(\frac{N_a}{n_i}\right),
\end{equation}
where $E_G$ is the energy gap, $n_{i}$ the intrinsic carriers density, and $kT/q$ the thermal voltage. Results are reported in \textbf{Figure \ref{fig:figure_3}b}: a limited difference is observed between the two combinations of metal and oxide, suggesting a similar quality of the interface between Cr/Co and gate oxide with respect to the case of Ti/Pd. Finally, we studied the impact of the thermal treatment on the electrical conduction of Co-FETs (\textbf{Figure \ref{fig:figure_3}c}). The sharp increase in transconductance ($g_m=d I_D/d V_{GS}$) observed in the annealed samples is consistent with the passivation of dangling bonds and reduction of oxide charges, and it indicates a decrease in the contact resistance of the devices. 

\section{FDSOI Nanowires with Ferromagnetic Cobalt Gates}
To demonstrate the co-integration of ferromagnetic gates with standard metallic control gates, an FDSOI nanowire device comprising two quantum dots has been fabricated (Figure \ref{fig:figure_1}). The top $<$100$>$ silicon layer facing the buried-oxide was selectively etched down to 18 nm and the nanowire patterned by e-beam lithography and reactive ion etching with SF$_6$/C$_4$F$_8$. The width of the nanowire was set to 50 nm to facilitate electrostatic and physical quantum confinement, whereas wider nanowires should be adopted for the control of separated corner dots with split gates, as successfully reported for quantum dots with holes \cite{maurand2016cmos}. Palladium plunger gates with a thin adhesion layer of titanium have been fabricated to control the chemical potential of the dots in the nanowire. They alternate with barrier gates made of chromium-cobalt, used to electrostatically tune depletion tunneling barriers for quantum confinement, and are insulated by ALD-Al$_2$O$_3$. Ohmic contacts have been fabricated with phosphorus-implanted n$^{++}$ wells ($N_d \approx 10^{19} \, cm^{-3}$) contacted with Ti/Pt to probe electron conduction through the structure from room temperature to millikelvin with minimal parasitic series resistance. To isolate the quantum dots from diffused dopants, two lateral Pd inversion gates extend the highly conductive implanted regions to the dots area, creating two continuous inversion layers. All layers have been fabricated with e-beam lithography, metal e-beam evaporation and lift-off process. The average pitch is 70 nm and a barrier gate 60 nm wide alternates with a pair of 35 nm Co gates; this arrangement creates varying stray fields in the region beneath the plunger gates, resulting in different Larmor frequencies in the two quantum dots (see subsection 3.2.1). It is worth mentioning that industrial processing relies on dry etching techniques, and Co can be etched using such methods \cite{kim2023high}, providing a promising solution for large-scale integration of nanomagnets.

\begin{figure}
\centering
  \includegraphics[width=0.8\linewidth]{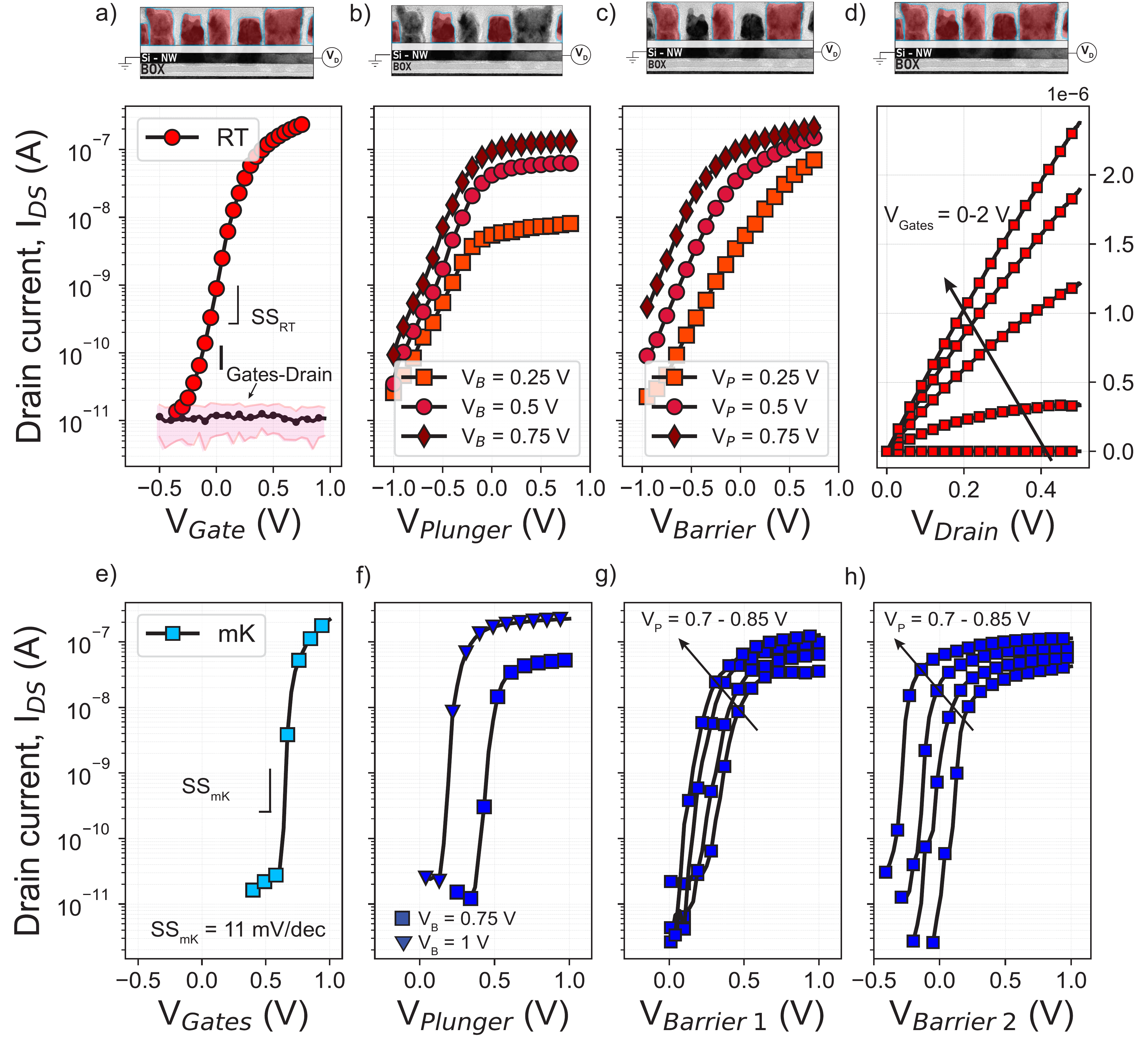}
  \caption{Electrical characterization of the fabricated FDSOI nanowire. Room temperature measurements (the gates activated in the voltage sweeps are colored in the TEM images): transcharacteristic curves for a) all gates with mean value and standard deviation of the leakage current from gates to channel (I$_{\mathrm{Gates-Drain}}$), b) Ti-Pd plunger gates at different barrier gates voltage, c) Cr-Co barrier gates at different plunger gates voltage, and d) bias curves for all gates above threshold. Cryogenic characterization of gates defining a quantum dot: transcharacteristic curves for e) all gates, f) one plunger gate, g-h) barrier gates at different plunger gate voltages.}
  \label{fig:figure_4}
\end{figure}

The independent electrical control of the two layers of gates at room temperature is reported in \textbf{Figure \ref{fig:figure_4} (a-d)}. Each layer exhibits pinch-off curves with subthreshold swing (SS$_{\mathrm{RT}}$) below 100 mV/dec and ohmic behaviour of the contacts. A threshold voltage mismatch ($\Delta V_{th} \approx 0.5 \; V$) is measured between the two layers, due to the interlayer oxide and difference in work-function. Finally, the linear threshold voltage dependence of barrier gates on the plunger’s bias (and vice versa) suggests a low inter-layers defectivity \cite{niebojewski2022specificities}, validating the quality of insulation between the metals. \textbf{Figure \ref{fig:figure_4} (e-h)} reports similar curves for a plunger gate and two barrier gates defining a quantum dot at cryogenic temperature (10 mK). Here, the discrepancy between the characteristics of the two barriers stems from the varying dimensions of the cobalt gates which result in a different capacitance to the semiconductor.

\subsection{Quantum Confinement with Cobalt Barrier Gates}
\begin{figure}
\centering
  \includegraphics[width=0.8\linewidth]{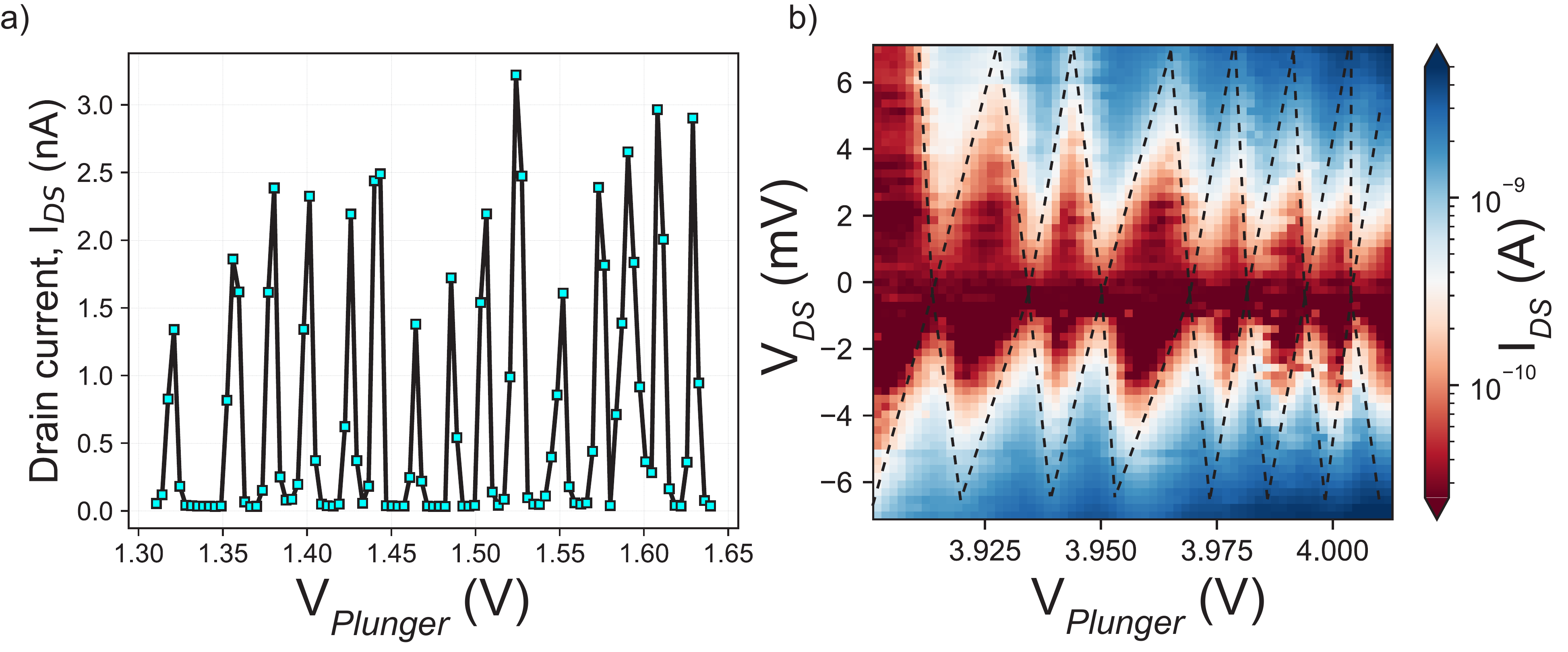}
  \caption{Quantum confinement induced by Co barrier gates at 10 mK. a) Coulomb blockade peaks for a quantum dot electrostatically induced by two ferromagnetic Co barrier gates. b) Coulomb blockade diamonds for the same dot at lower barriers voltage.}
  \label{fig:figure_5}
\end{figure}

Formation of quantum dots in the many-electron regime has been investigated with DC Coulomb blockade spectroscopy at 11 mK. \textbf{Figure \ref{fig:figure_5}} displays periodic Coulomb oscillations and diamonds for a quantum dot, obtained by setting the ferromagnetic barrier gates to values below their threshold voltage and sweeping the plunger gate voltage until reaching strong inversion. A dot capacitance of $C_{dot}\approx$ 27 aF is estimated from the diamonds, giving an equivalent dot radius of $r_{dot} \approx \; $35 nm in a self-capacitance disc model ($C_{dot}=8\epsilon_{eq} \epsilon_0 r_{dot}$, where $\epsilon_{eq}=(\epsilon_{Si}+\epsilon_{Al_2O_3})/2$),  in agreement with the dimension of the nanowire. A lever arm of $\alpha_P \approx$ 0.25 eV/V is extracted for plunger gates on 9 nm of ALD-Al$_2$O$_3$ deposited as the gate oxide. It is worth to mention that a factor of 2x in the lever arm value (and therefore in the capacitive coupling to the channel) for both layers and better electrostatics could be achieved by employing a thinner thermal gate oxide (dry-SiO$_2$) without any additional interlayer ALD-dielectric, as in the case of quantum dots featuring oxidized aluminium gates on SiO$_2$ \cite{saraiva2022materials}. Due to the reduced density of interface traps, this approach would also be advantageous in terms of charge and current noise, which are critical metrics for achieving a stable charge configuration in spin qubit operation \cite{elsayed2024low}.

\subsection{Nanogates Magnetization Pattern}
\begin{figure}
\centering
  \includegraphics[width=\linewidth]{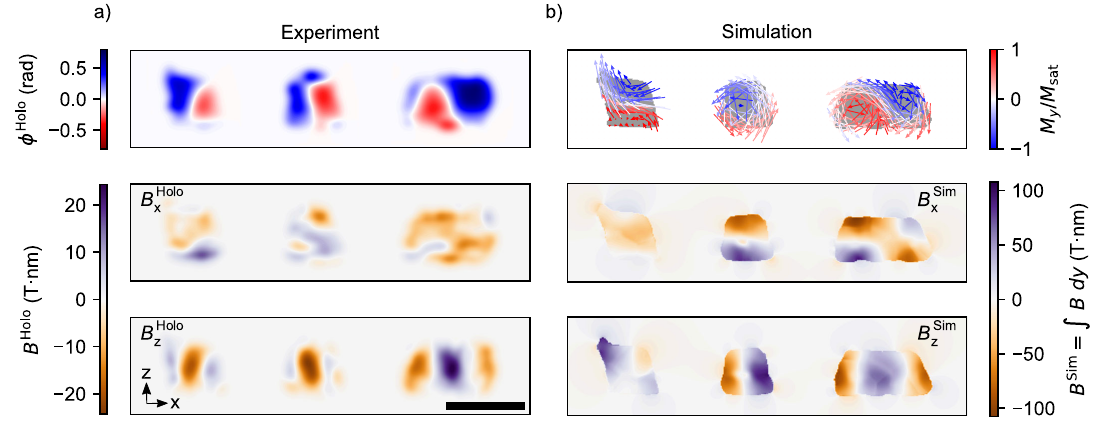}
  \caption{Electron holography and micromagnetic simulations comparison. a) Electron holography measurement with magnetic phase ($\phi^{\mathrm{Holo}}$) (upper panel) and $x$-$z$-component (middle and bottom panel, respectively) of the magnetic induction integrated over the sample thickness ($\boldsymbol{B^{\mathrm{Holo}}}$). b) Micromagnetic simulation of Co nanogates. The upper panel shows the magnetization pattern in the center of the lamella, while the middle and bottom panel the $x$- and $z$-component of the magnetic induction integrated over $y$ ($\boldsymbol{B^{\mathrm{sim}}}$). The scale bar is 50 nm. }
  \label{fig:figure_6}
\end{figure}

To gain insight into the magnetic domain structures of the Co control gates, electron holography experiments \cite{thomas2008electron} have been performed. A thin ($t\approx40$ nm) sample cross-section was prepared using focused ion beam (FIB) to investigate three Co gates with different widths and constant thickness. Measurements were performed at null external magnetic field ($B_{\mathrm{ext}}$) after magnetizing the sample along the in-plane $\hat{x}$ direction (i.e., the direction of the SOI nanowire). The magnetic phase ($\phi^{\mathrm{Holo}}$) of an electron travelling through the sample along the $\hat{y}$ direction is given by the magnetic vector potential \textbf{A} according to:
\begin{equation}
    \phi^{\mathrm{Holo}}(x,z) = - \frac{e}{\hbar}\int_{-\infty}^{+\infty} A_y(\textbf{r})dy,
\end{equation}
where \textbf{r} is the position vector, $A_y$ the $y$-component of the magnetic vector potential, $e$ the elementary charge, and $\hbar$ the reduced Planck's constant. The electron holography measurement allows mapping the in-plane components of magnetic induction (i.e., $\hat{x}, \hat{z}$) integrated along the thickness of the sample in the Coulomb gauge ($\textbf{B}(\textbf{r}) = \nabla \times \textbf{A}(\textbf{r})$, where $\nabla$ is the nabla operator and $\times$ the cross product, see SI for details). Mapping of the magnetic phase and the magnetic induction ($\boldsymbol{B^{\mathrm{Holo}}}$), integrated over the out-of-plane $\hat{y}$ direction, are reported in \textbf{Figure \ref{fig:figure_6}}. As expected, the measured remanent magnetization does not align with the magnetization axis, and an intercoupling between two magnetic domains is observed in the larger Co pad. To investigate the role of polycrystallinity,  micromagnetic simulations with a 20 nm sized crystallite random pattern  were performed. The simulation is run on a mesh of polyhedra defined using the method of Voronoi tesselation and assigns random crystal directions to each polyhedron, mimicking the non-textured polycrystalline structure shown by XRD and TEM. \textbf{Figure~\ref{fig:figure_6}b} shows one simulation outcome with a magnetization pattern that leads to a $\boldsymbol{B^{\mathrm{Sim}}}$ closely resembling the holography experiment, at least for the middle and rightmost magnetic pad. The vortex-like structures visible in the middle and rightmost magnetic pads minimize the stray field, but the overall complicated magnetic distribution suggests that the crystallite texture and fabrication-induced defects play a major role in determining the local magnetization orientation. Indeed,  simulations which differ only by a different crystallite pattern do not lead to similar results (see SI). This suggests investigating a lower bound for the external magnetic field $B_{\mathrm{ext}}$, knowing that the requirements for an effective control of spin states through EDSR (large driving gradients, small dephasing gradients and single qubit addressability~\cite{yoneda2015robust}) demand uniform magnetization along the nanowire axis~\cite{bersano2023quantum}. 

\subsubsection{EDSR Driving by Co Nanogates}
\begin{figure}
\centering
  \includegraphics[width=0.7\linewidth]{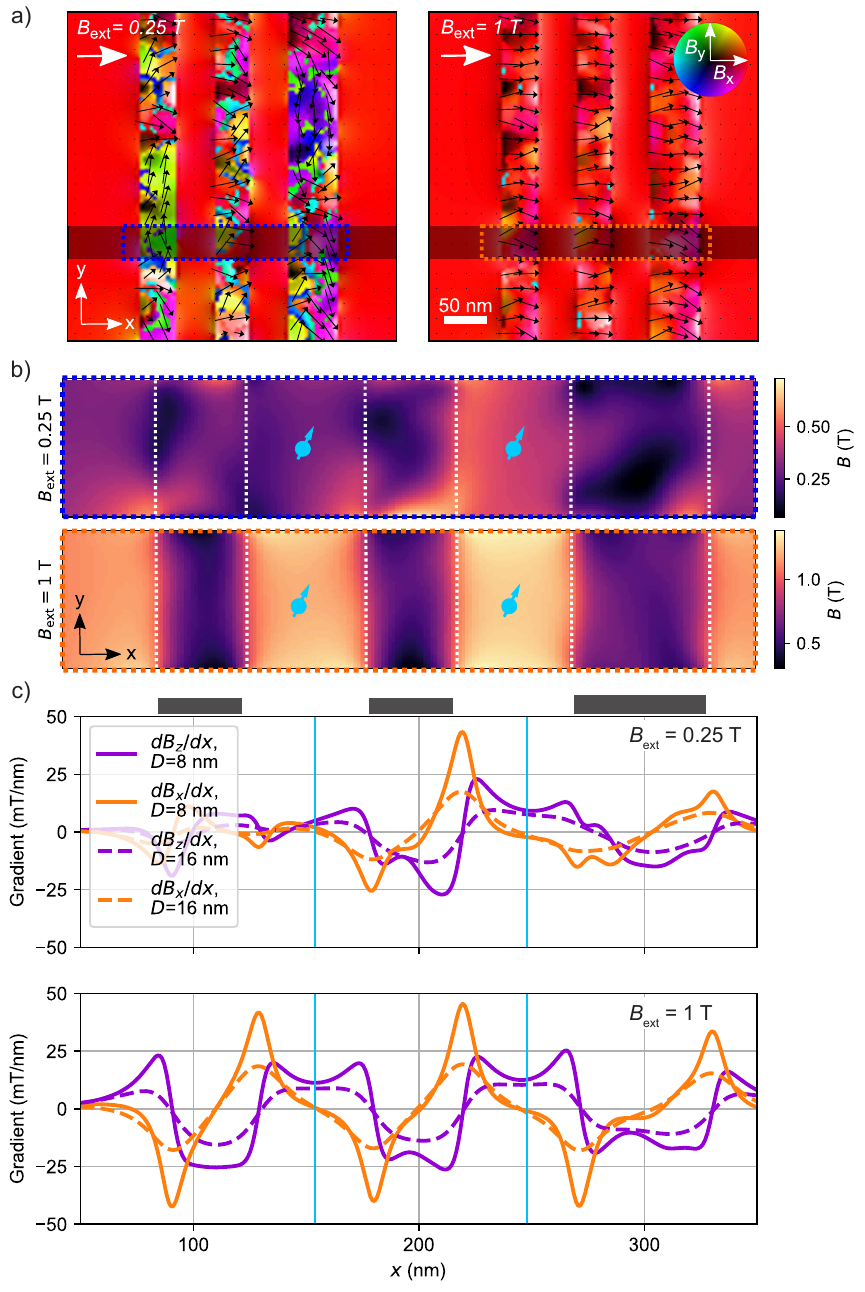}
  \caption{Micromagnetic simulation of Co nanogates. a) Total in-plane magnetic field ($\boldsymbol{B^{\mathrm{Sim}}}$) and magnetization pattern (black arrows) at different external magnetic fields ($B_{\mathrm{ext}}$). The nanowire position is shown as an overlaid black rectangle. The dotted rectangles show the position of the areas plotted in the figure below. b) Stray field in the qubit plane (i.e., inside the SOI nanowire) at different external magnetic fields. The positions of the qubits are shown by blue circles, while the dotted lines represent the projected edges of the magnetic nanogates wrapping the SOI nanowire. c) Driving and dephasing gradients at low and high field in the plane of the qubits for two different magnet-to-dot distances relative to Co deposited as the first and second metallization layer, respectively. Blue lines mark the position of the qubits, while grey rectangles indicate the position of the Co gates.}
  \label{fig:figure_7}
\end{figure}

The minimum value of $B_{\mathrm{ext}}$ to enforce uniform magnetization is investigated by micromagnetic simulations, where the shape of the nanogates wrapping the Si nanowire is taken into account, as well as the polycrystalline structure of Co (see SI). 
\textbf{Figure \ref{fig:figure_7}a} shows that  $B_{\mathrm{ext}}=1$\,T ensures an almost homogeneous magnetization within the nanogates, whereas at $B_{\mathrm{ext}}=0.25$\,T the magnetic shape anisotropy and magnetocrystalline anisotropy dominate the magnetization pattern orientation. This leads to a non-uniform stray field in the qubit plane for $B_{\mathrm{ext}}< 1$ T, as shown in \textbf{Figure \ref{fig:figure_7}b}, with the result of an impredictable addressability (i.e., difference in Larmor frequency) of the two adjacent qubits. Assuming sufficiently high fields, a strong magnetic field gradient along the direction normal to the quantization axis ($\hat{x}$) can be created inside the nanowire due to the reduced distance between the ferromagnets and the semiconductor. For qubit operation, the driving gradient is $dB_z/dx$, while dephasing is mainly caused by $dB_x/dx$, since the field symmetry induced by the wrapping gates makes $dB_y/dx$ negligible at the qubit plane. We report the simulated gradients generated by the fabricated Co nanogates at low and high field in \textbf{Figure \ref{fig:figure_7}c}, considering a magnet-to-dot distance ($D$) defined by the gate oxide (with $D=8$ nm) or the gate and interlayer oxide (with $D=16$ nm), depending on whether the Co nanogates are patterned as the first or second metallization layer. At $B_{\mathrm{ext}}=0.25$ T, the unsaturated magnetization along $\hat{x}$ results in a distorted gradient shape. However, at $B_{\mathrm{ext}}=1$ T, driving gradients exceeding 10 mT/nm are achieved in both configurations, with the decoherence gradient being minimized at the location of the quantum dots. In this experimental condition, the predicted Larmor frequency difference ($\Delta f_L$, assuming a g-factor $g=2$) between the two qubits is $3.78$ GHz and $2.77$ GHz for the two gate-stack configurations, respectively. This is beneficial in terms of qubit fidelity, since a fast-driving region coincides with a dephasing protected spot along the main displacement direction, with $\Delta f_L$ more than two times larger than the estimated Rabi frequency \cite{bersano2023quantum}. Based on these micromagnetic simulation results, we conclude that the fabricated Co nanogates are suitable for EDSR drive of spin qubits at high Zeeman fields only ($B_{\mathrm{ext}}\geq1$\,T), where homogeneous magnetization is ensured along the hard magnetic axis of the nanostructures.\\
These considerations, based on numerical simulations, are independent of the dielectric used, assuming a fixed oxide thickness and a consistent distance between the ferromagnets and quantum dots. In the proof-of-concept devices presented in this study, we observed significant charge noise in gate-stack utilizing exclusively Pd gates, as well as in configurations combining Pd and Co control gates. This charge noise, which we attribute to the ALD-Al$_2$O$_3$/Si interface, impeded the precise tuning of double-dot charge configurations required to map a Rabi Chevron pattern via EDSR and Pauli spin blockade. As a result, the demonstration of electron spin qubit operation within the proposed scalable EDSR architecture is deferred to future studies utilizing high-temperature thermal SiO$_2$ in place of ALD-based dielectrics.

\section{Conclusion}
In this study we analyzed the integration of Co nanogates for electrical and magnetic manipulation of quantum dots in Si and validated its compatibility with FEOL CMOS processes. We examined the chemistry and morphology of e-beam evaporated thin Co films using ARXPS, GIXRD, and EDX-TEM. Our observations revealed interlayer diffusion of the adhesion layer (Cr) into Co after thermal treatment in forming gas at 300\;°C, along with limited surface oxidation in samples that were carefully cooled in nitrogen. XPS spectra confirmed the formation of CoO/Co$_2$O$_3$/Co$_3$O$_4$ and Co(OH$_2$) at the surface of the sample. Under our process conditions, the partial oxidation was limited to 4 nm and did not impact the magnetic properties of the underlying Co, as measured by VSM. Electron diffraction patterns indicate the formation of an \textit{fcc} phase for annealing temperatures above 300\;°C, and limited evolution of the grain size was measured within the temperature range used in this work. We compared the work function and charge traps of Ti-Pd and Cr-Co gate stacks by performing high-low frequency characterization of MOS capacitors, noting a 15$\%$ increase in effective work function of the ferromagnetic gate-stack and a reduction of charge traps to $10^{10}$ cm$^{-3}$ when thermally annealed in forming gas at 300\;°C for 10 minutes. No relevant difference of interface traps was measured between Ti/Pd and Cr/Co gate stacks. To validate the process, we fabricated an FDSOI nanowire device featuring Pd and Co control gates for the electrostatic control of quantum dots. Electrical characterizations at room temperature and 10 mK confirmed accurate field-effect control of the charge density in SOI, and stable quantum confinement was observed at cryogenic temperature for Co gates biased below threshold. Electron holography measurements coupled with micromagnetic simulations suggest that local magnetic texture inside the ferromagnetic gates is mostly controlled by the crystallite structure and fabrication-induced defects, both dependent on thermal treatments and material morphology. Micromagnetic simulations mimicking the non-textured polycrystalline structure of Co measured by XRD and TEM have been used to determine a lower bound for the external magnetic field to ensure homogeneous magnetization along the hard axis of the ferromagnetic gates. Simulation results show that for magnetic fields lower than 1 T, the magnetic shape and magnetocrystalline anisotropy lead to a non-uniform stray field in the plane of the quantum dots which ultimately degrades the control over driving and dephasing gradients, thus establishing a minimum operation condition for the EDSR drive. These findings demonstrate the feasibility of FEOL/MOL-compatible integration of Co in silicon-based quantum devices, with emphasis on ferromagnetic control gates suitable for both electrostatic control and EDSR driving of electron spin qubits.

\section{Experimental Section}
\textit{Samples Fabrication}: The SOI substrates were provided by Soitec and thinned down to 18 nm by several cycles of dry oxidation in a Centrotherm furnace and wet etching in buffered HF. The SOI nanowires and all metal layers were patterned by electron beam lithography using a Raith EPBG5000+ system. A commercial hydrogen silsesquioxane solution (HSQ XR-1541 2$\%$) was spin-coated on the SOI sample at 7000 rpm to achieve a uniform thickness of approximately 50 nm. The resist was exposed to a dose of $2000$ $\mu$C/cm$^2$ and developed in 25$\%$ TMAH at room temperature for 50 s to define a hard mask for the silicon nanowire. The nanowire was then anisotropically etched in an inductively coupled deep reactive ion etching system (Alcatel AMS200) using an SOI-optimized continous etch recipe comprising SF$_6$ and C$_4$F$_8$, with the selectivity to HSQ estimated to be 20:1. The hard mask was then removed by a $30$ s dip in HF $1\%$, given the much higher etch rate compared to the thermal SiO$_2$ forming the buried oxide. A 9 nm Al$_2$O$_3$ gate oxide was deposited by ALD at 300\;°C using trimethylalluminum and H$_2$O precursors in a Beneq TFS200 system. Both metallic layers were patterned by lift-off using a single layer of PMMA 950K with an approximate thickness of 90 nm. All metals were deposited via e-beam evaporation: 3 nm for adhesion layers (Ti, Cr) and 33 nm for Pd and Co. The evaporation rate of Co was set to 4 $\AA$/s at a pressure lower than 1.5$\cdot 10^{-6}$ mbar, and the temperature reached during deposition was monitored to remain below 80 °C. Low power oxygen plasma was used to remove any residues of polymers before metals evaporation. A 7 nm ALD-deposited aluminum oxide layer served as the interlayer insulator. Finally, all contacts were opened by maskless laser lithography and wet etching in BHF to access metallic pads for wire-bonding.
The TEM lamellae were prepared by Ga FIB and stored under vacuum. \\ 
\textit{Material Characterization}: GIXRD analysis was performed on a Malvern Panalytical Empyrean thin films diffractometer, equipped with a monochromated Cu K$\alpha$ source and a PIXcel$^{3D}$ detector. XPS and ARXPS analyses were performed using a Kratos Axis Supra equipped with a monochromated Al K$\alpha$ source. SEM EDX imaging was conducted using a Zeiss Merlin SEM equipped with a X-Max 50 mm$^2$ detector from Oxford Instruments. TEM imaging, EDX and SAED measurements were performed using a FEI Tanos F200S microscope, operated at 200 kV and equipped with a Thermo Fisher CETA camera. The VSM measurements were performed with a Lake Shore Cryotronics 7300 Series Vibrating Sample Magnetometer. Hysteresis loops were measured by applying the external field parallel to the surface of the thin film (i.e. within the easy plane set by shape anisotropy). Electron holography measurements were performed on a FEI Titan Themis microscope operated at 300 kV, equipped with an electron biprism and a Gatan K3 camera; $\pi$-phase shifting off-axis holography method was used \cite{boureau_off-axis_2018} and electrostatic contribution was removed to access the magnetic contribution.\\
\textit{Electrical Measurements}: Capacitance characterization and room temperature field-effect measurements were performed on a probe station using a Keithley 4200A-SCS parameter analyzer equipped with CVUs and pre-amplifiers. All capacitance values reported in this work have been corrected by the series resistance $R_S = \frac{(G_{am}/2\pi f C_{am})^2}{[1+(\frac{G_{am}}{2\pi f C_{am}})^2]G_{am}}$ according to $C=\frac{(G_{am}^2 + (2\pi fC_{am})^2)C_{am}}{[G_{am}-(G_{am}^2+(2\pi fC_{am})^2)R_S]^2+(2\pi fC_{am})^2}$, where $C_{am}$ and $G_{am}$ are, respectively, the as-measured capacitance and conductance using the parallel model ($C_p-G_p$), and $f$ the test frequency \cite{nicollian2002mos}. Cryogenic measurements were taken in a Bluefors dilution refrigerator with a base temperature of 10 mK, equipped with digital multimeter, current-to-voltage converter and a multi-channel ultra-low-noise DAC.  \\
\textit{Micromagnetic Simulations}: All micromagnetic simulations have been performed with the software MuMax3~\cite{vansteenkiste2014design}. For the simulations of the magnetic lamella, a cell size of 1 nm, a saturation magnetization ($M_{\mathrm{sat}}$) of 1.44 MA/m (as measured by vibrating sample magnetometry) and an exchange stiffness ($A$) of 25 pJ/m have been used. The polycrystalline structure has been simulated by using the function named \verb|ext_make3dgrains|, with a mean grain size of 20 nm and an uniaxial magnetocrystalline anisotropy of $K_{U1}=400$\,KJ/$\mathrm{m}^3$ and $K_{U2}=180$\,KJ/$\mathrm{m}^3$. The nanogates lamella shape for the simulations has been constructed by reproducing the magnet cross-section from the TEM images to the simulation mesh and extruding the cross-section along the out-of-plane direction. The thickness of the magnets has been set to 40 nm, as estimated from the lamella thickness. For the simulations of the Co nanogates the same $M_{\mathrm{sat}}$, $A$ and cell size have been used, but the magnetocrystalline anisotropy constant temperature dependence has been taken into account by setting $K_{U1}=650$\,KJ/$\mathrm{m}^3$ and $K_{U2}=180$\,KJ/$\mathrm{m}^3$, since spin qubits are operated close to 0\,K. 

\bigskip
\textbf{Supporting Information} \par 
Supporting Information for XRD and XPS analysis, electron holography measurements, and micromagnetic simulations is available.

\bigskip
\textbf{Acknowledgements} \par 
The authors acknowledge the Center of Micro-NanoTechnology (CMi) and the Interdisciplinary Center for Electron Microscopy (CIME) of EPFL for their support during the development of the fabrication process and materials characterization, and the X-Ray Diffraction and Surface Analytics Facility for XPS measurements. A special gratitude goes to Dr. Rolf Allenspach (IBM Research-Zurich) for his valuable insights on the micromagnetic simulations, as well as to Dr. Floris Braakman, Liza Žaper, and prof. Martino Poggio from the University of Basel for their insightful discussions regarding the presented architecture. This work was supported as a part of NCCR SPIN, a National Centre of Competence in Research, funded by the Swiss National Science Foundation (grant number 225153).
\medskip

%
\bibliographystyle{MSP}


\bibliography{references}

\begin{thebibliography}{10}
\providecommand{\url}[1]{\texttt{#1}}
\providecommand{\urlprefix}{URL }

\bibitem{finocchio2023roadmap}
G.~Finocchio, J.~A.~C. Incorvia, J.~S. Friedman, Q.~Yang, A.~Giordano, J.~Grollier, H.~Yang, F.~Ciubotaru, A.~Chumak, A.~Naeemi, et~al.,
\newblock \emph{Nano Futures} \textbf{2023}, \emph{8}, 1 012001.

\bibitem{leonard2022shape}
T.~Leonard, S.~Liu, M.~Alamdar, H.~Jin, C.~Cui, O.~G. Akinola, L.~Xue, T.~P. Xiao, J.~S. Friedman, M.~J. Marinella, et~al.,
\newblock \emph{Advanced Electronic Materials} \textbf{2022}, \emph{8}, 12 2200563.

\bibitem{jung2022crossbar}
S.~Jung, H.~Lee, S.~Myung, H.~Kim, S.~K. Yoon, S.-W. Kwon, Y.~Ju, M.~Kim, W.~Yi, S.~Han, et~al.,
\newblock \emph{Nature} \textbf{2022}, \emph{601}, 7892 211.

\bibitem{watson2018programmable}
T.~Watson, S.~Philips, E.~Kawakami, D.~Ward, P.~Scarlino, M.~Veldhorst, D.~Savage, M.~Lagally, M.~Friesen, S.~Coppersmith, et~al.,
\newblock \emph{Nature} \textbf{2018}, \emph{555}, 7698 633.

\bibitem{pioro2008electrically}
M.~Pioro-Ladrière, T.~Obata, Y.~Tokura, Y.-S. Shin, T.~Kubo, K.~Yoshida, T.~Taniyama, S.~Tarucha,
\newblock \emph{Nature Physics} \textbf{2008}, \emph{4}, 10 776.

\bibitem{yoneda2018quantum}
J.~Yoneda, K.~Takeda, T.~Otsuka, T.~Nakajima, M.~R. Delbecq, G.~Allison, T.~Honda, T.~Kodera, S.~Oda, Y.~Hoshi, et~al.,
\newblock \emph{Nature Nanotechnology} \textbf{2018}, \emph{13}, 2 102.

\bibitem{kawakami2014electrical}
E.~Kawakami, P.~Scarlino, D.~R. Ward, F.~Braakman, D.~Savage, M.~Lagally, M.~Friesen, S.~N. Coppersmith, M.~A. Eriksson, L.~Vandersypen,
\newblock \emph{Nature Nanotechnology} \textbf{2014}, \emph{9}, 9 666.

\bibitem{leon2020coherent}
R.~Leon, C.~H. Yang, J.~Hwang, J.~C. Lemyre, T.~Tanttu, W.~Huang, K.~W. Chan, K.~Tan, F.~Hudson, K.~Itoh, et~al.,
\newblock \emph{Nature Communications} \textbf{2020}, \emph{11}, 1 797.

\bibitem{zhang2021controlling}
X.~Zhang, Y.~Zhou, R.-Z. Hu, R.-L. Ma, M.~Ni, K.~Wang, G.~Luo, G.~Cao, G.-L. Wang, P.~Huang, et~al.,
\newblock \emph{Physical Review Applied} \textbf{2021}, \emph{15}, 4 044042.

\bibitem{philips2022universal}
S.~G. Philips, M.~T. Madzik, S.~V. Amitonov, S.~L. de~Snoo, M.~Russ, N.~Kalhor, C.~Volk, W.~I. Lawrie, D.~Brousse, L.~Tryputen, et~al.,
\newblock \emph{Nature} \textbf{2022}, \emph{609}, 7929 919.

\bibitem{yoneda2015robust}
J.~Yoneda, T.~Otsuka, T.~Takakura, M.~Pioro-Ladri{\`e}re, R.~Brunner, H.~Lu, T.~Nakajima, T.~Obata, A.~Noiri, C.~J. Palmstr{\o}m, et~al.,
\newblock \emph{Applied Physics Express} \textbf{2015}, \emph{8}, 8 084401.

\bibitem{bersano2023quantum}
F.~Bersano, M.~Aldeghi, E.~Collette, M.~Ghini, F.~De~Palma, F.~Oppliger, P.~Scarlino, F.~Braakman, M.~Poggio, H.~Riel, et~al.,
\newblock In \emph{2023 IEEE Symposium on VLSI Technology and Circuits (VLSI Technology and Circuits)}. IEEE, \textbf{2023} 1--2.

\bibitem{sabbagh2019quantum}
D.~Sabbagh, N.~Thomas, J.~Torres, R.~Pillarisetty, P.~Amin, H.~George, K.~Singh, A.~Budrevich, M.~Robinson, D.~Merrill, et~al.,
\newblock \emph{Physical Review Applied} \textbf{2019}, \emph{12}, 1 014013.

\bibitem{veldhorst2017silicon}
M.~Veldhorst, H.~Eenink, C.-H. Yang, A.~S. Dzurak,
\newblock \emph{Nature Communications} \textbf{2017}, \emph{8}, 1 1766.

\bibitem{gonzalez2021scaling}
M.~Gonzalez-Zalba, S.~De~Franceschi, E.~Charbon, T.~Meunier, M.~Vinet, A.~Dzurak,
\newblock \emph{Nature Electronics} \textbf{2021}, \emph{4}, 12 872.

\bibitem{zwerver2022qubits}
A.~Zwerver, T.~Kr{\"a}henmann, T.~Watson, L.~Lampert, H.~C. George, R.~Pillarisetty, S.~Bojarski, P.~Amin, S.~Amitonov, J.~Boter, et~al.,
\newblock \emph{Nature Electronics} \textbf{2022}, \emph{5}, 3 184.

\bibitem{klemt2023electrical}
B.~Klemt, V.~Elhomsy, M.~Nurizzo, P.~Hamonic, B.~Martinez, B.~Cardoso~Paz, C.~Spence, M.~C. Dartiailh, B.~Jadot, E.~Chanrion, et~al.,
\newblock \emph{npj Quantum Information} \textbf{2023}, \emph{9}, 1 107.

\bibitem{koch2024industrial300mmwaferprocessed}
T.~Koch, C.~Godfrin, V.~Adam, J.~Ferrero, D.~Schroller, N.~Glaeser, S.~Kubicek, R.~Li, R.~Loo, S.~Massar, G.~Simion, D.~Wan, K.~D. Greve, W.~Wernsdorfer \textbf{2024}, \emph{(Preprint)}, arXiv: 2409.12731.

\bibitem{aldeghi2023modular}
M.~Aldeghi, R.~Allenspach, G.~Salis,
\newblock \emph{Applied Physics Letters} \textbf{2023}, \emph{122}, 13 134003.

\bibitem{loss1998quantum}
D.~Loss, D.~P. DiVincenzo,
\newblock \emph{Physical Review A} \textbf{1998}, \emph{57}, 1 120.

\bibitem{elzerman2004single}
J.~Elzerman, R.~Hanson, L.~Willems~van Beveren, B.~Witkamp, L.~Vandersypen, L.~P. Kouwenhoven,
\newblock \emph{Nature} \textbf{2004}, \emph{430}, 6998 431.

\bibitem{rolf_gian_patent}
R.~Allenspach, G.~Salis,
\newblock \emph{US patent} 7,336,515.

\bibitem{veldhorst2014addressable}
M.~Veldhorst, J.~Hwang, C.~Yang, A.~Leenstra, B.~de~Ronde, J.~Dehollain, J.~Muhonen, F.~Hudson, K.~M. Itoh, A.~t. Morello, et~al.,
\newblock \emph{Nature Nanotechnology} \textbf{2014}, \emph{9}, 12 981.

\bibitem{wang2022ultrafast}
K.~Wang, G.~Xu, F.~Gao, H.~Liu, R.-L. Ma, X.~Zhang, Z.~Wang, G.~Cao, T.~Wang, J.-J. Zhang, et~al.,
\newblock \emph{Nature Communications} \textbf{2022}, \emph{13}, 1 206.

\bibitem{camenzind2022hole}
L.~C. Camenzind, S.~Geyer, A.~Fuhrer, R.~J. Warburton, D.~M. Zumb{\"u}hl, A.~V. Kuhlmann,
\newblock \emph{Nature Electronics} \textbf{2022}, \emph{5}, 3 178.

\bibitem{gilbert2023demand}
W.~Gilbert, T.~Tanttu, W.~H. Lim, M.~Feng, J.~Y. Huang, J.~D. Cifuentes, S.~Serrano, P.~Y. Mai, R.~C. Leon, C.~C. Escott, et~al.,
\newblock \emph{Nature Nanotechnology} \textbf{2023}, \emph{18}, 2 131.

\bibitem{liu2024progress}
Y.~Liu, S.~Guan, J.-W. Luo, S.-S. Li,
\newblock \emph{Advanced Functional Materials} \textbf{2024}, 2304725.

\bibitem{forster2015electric}
F.~Forster, M.~M{\"u}hlbacher, D.~Schuh, W.~Wegscheider, S.~Ludwig,
\newblock \emph{Physical Review B} \textbf{2015}, \emph{91}, 19 195417.

\bibitem{tadokoro2021designs}
M.~Tadokoro, T.~Nakajima, T.~Kobayashi, K.~Takeda, A.~Noiri, K.~Tomari, J.~Yoneda, S.~Tarucha, T.~Kodera,
\newblock \emph{Scientific Reports} \textbf{2021}, \emph{11}, 1 19406.

\bibitem{bordoloi2022spin}
A.~Bordoloi, V.~Zannier, L.~Sorba, C.~Sch{\"o}nenberger, A.~Baumgartner,
\newblock \emph{Nature} \textbf{2022}, \emph{612}, 7940 454.

\bibitem{borsoi2024shared}
F.~Borsoi, N.~W. Hendrickx, V.~John, M.~Meyer, S.~Motz, F.~van Riggelen, A.~Sammak, S.~L. de~Snoo, G.~Scappucci, M.~Veldhorst,
\newblock \emph{Nature Nanotechnology} \textbf{2024}, \emph{19}, 1 21.

\bibitem{lawrie2020quantum}
W.~Lawrie, H.~Eenink, N.~Hendrickx, J.~Boter, L.~Petit, S.~Amitonov, M.~Lodari, B.~Paquelet~Wuetz, C.~Volk, S.~Philips, et~al.,
\newblock \emph{Applied Physics Letters} \textbf{2020}, \emph{116}, 8.

\bibitem{cullity2011introduction}
B.~D. Cullity, C.~D. Graham,
\newblock \emph{Introduction to magnetic materials},
\newblock John Wiley \& Sons, \textbf{2011}.

\bibitem{hubert2008magnetic}
A.~Hubert, R.~Sch{\"a}fer,
\newblock \emph{Magnetic Domains: the Analysis of Magnetic Microstructures},
\newblock Springer Science \& Business Media, \textbf{2008}.

\bibitem{istratov2000iron}
A.~Istratov, H.~Hieslmair, E.~Weber,
\newblock \emph{Applied Physics A} \textbf{2000}, \emph{70} 489.

\bibitem{narasimha20177nm}
S.~Narasimha, B.~Jagannathan, A.~Ogino, D.~Jaeger, B.~Greene, C.~Sheraw, K.~Zhao, B.~Haran, U.~Kwon, A.~Mahalingam, et~al.,
\newblock In \emph{2017 IEEE International Electron Devices Meeting (IEDM)}. IEEE, \textbf{2017} 29--5.

\bibitem{ichimori2002fully}
T.~Ichimori, N.~Hirashita,
\newblock \emph{IEEE Transactions on Electron Devices} \textbf{2002}, \emph{49}, 12 2296.

\bibitem{kamineni2016tungsten}
V.~Kamineni, M.~Raymond, S.~Siddiqui, F.~Mont, S.~Tsai, C.~Niu, A.~Labonte, C.~Labelle, S.~Fan, B.~Peethala, et~al.,
\newblock In \emph{2016 IEEE International Interconnect Technology Conference/Advanced Metallization Conference (IITC/AMC)}. IEEE, \textbf{2016} 105--107.

\bibitem{auth201710nm}
C.~Auth, A.~Aliyarukunju, M.~Asoro, D.~Bergstrom, V.~Bhagwat, J.~Birdsall, N.~Bisnik, M.~Buehler, V.~Chikarmane, G.~Ding, et~al.,
\newblock In \emph{2017 IEEE International Electron Devices Meeting (IEDM)}. IEEE, \textbf{2017} 29--1.

\bibitem{el2010inter}
T.~El~Asri, M.~Raissi, S.~Vizzini, A.~El~Maachi, E.~Ameziane, F.~A. d’Avitaya, J.-L. Lazzari, C.~Coudreau, H.~Oughaddou, B.~Aufray, et~al.,
\newblock \emph{Applied Surface Science} \textbf{2010}, \emph{256}, 9 2731.

\bibitem{dash2008near}
S.~Dash, D.~Goll, H.~Carstanjen,
\newblock \emph{Applied Physics A} \textbf{2008}, \emph{91} 379.

\bibitem{peng2006adsorption}
G.~Peng, A.~Huan, E.~Tok, Y.~Feng,
\newblock \emph{Physical Review B} \textbf{2006}, \emph{74}, 19 195335.

\bibitem{vcechal2008morphology}
J.~{\v{C}}echal, J.~Luksch, K.~Ko{\v{n}}{\'a}kov{\'a}, M.~Urb{\'a}nek, E.~Brandejsov{\'a}, T.~{\v{S}}ikola,
\newblock \emph{Surface Science} \textbf{2008}, \emph{602}, 15 2693.

\bibitem{kim2022robust}
C.~Kim, G.~Kang, Y.~Jung, J.-Y. Kim, G.-B. Lee, D.~Hong, Y.~Lee, S.-G. Hwang, I.-H. Jung, Y.-C. Joo,
\newblock \emph{Scientific Reports} \textbf{2022}, \emph{12}, 1 12291.

\bibitem{smardz1992oxidation}
L.~Smardz, U.~K{\"o}bler, W.~Zinn,
\newblock \emph{Journal of Applied Physics} \textbf{1992}, \emph{71}, 10 5199.

\bibitem{tompkins1981oxidation}
H.~Tompkins, J.~Augis,
\newblock \emph{Oxidation of Metals} \textbf{1981}, \emph{16} 355.

\bibitem{fontaina2014room}
N.~Fonta{\'i}{\~n}a-Troiti{\~n}o, S.~Li{\'e}bana-Vi{\~n}as, B.~Rodr{\'i}guez-Gonz{\'a}lez, Z.-A. Li, M.~Spasova, M.~Farle, V.~Salgueiri{\~n}o,
\newblock \emph{Nano Letters} \textbf{2014}, \emph{14}, 2 640.

\bibitem{chastain1992handbook}
J.~Moulder, J.~Chastain,
\newblock \emph{Handbook of X-ray Photoelectron Spectroscopy: A Reference Book of Standard Spectra for Identification and Interpretation of XPS Data},
\newblock Physical Electronics Division, Perkin-Elmer Corporation, \textbf{1992}.

\bibitem{jergel2009annealing}
M.~Jergel, I.~Cheshko, Y.~Halahovets, P.~{\v{S}}iffalovi{\v{c}}, I.~Mat'Ko, R.~Senderak, S.~Protsenko, E.~Majkova, {\v{S}}.~Luby,
\newblock \emph{Journal of Physics D: Applied Physics} \textbf{2009}, \emph{42}, 13 135406.

\bibitem{klinger2015crystallographic}
M.~Klinger, A.~J{\"a}ger,
\newblock \emph{Journal of Applied Crystallography} \textbf{2015}, \emph{48}, 6 2012.

\bibitem{klinger2017more}
M.~Klinger,
\newblock \emph{Journal of Applied Crystallography} \textbf{2017}, \emph{50}, 4 1226.

\bibitem{boter2020scaling}
J.~Boter \textbf{2020}, \emph{PhD Thesis} Delft University of Technology.

\bibitem{brauns2018palladium}
M.~Brauns, S.~V. Amitonov, P.-C. Spruijtenburg, F.~A. Zwanenburg,
\newblock \emph{Scientific Reports} \textbf{2018}, \emph{8}, 1 5690.

\bibitem{nicollian2002mos}
E.~H. Nicollian, J.~R. Brews,
\newblock \emph{MOS (Metal Oxide Semiconductor) Physics and Technology},
\newblock John Wiley \& Sons, \textbf{2002}.

\bibitem{gu2006effective}
D.~Gu, S.~K. Dey, P.~Majhi,
\newblock \emph{Applied Physics Letters} \textbf{2006}, \emph{89}, 8.

\bibitem{yeo2002effects}
Y.-C. Yeo, P.~Ranade, T.-J. King, C.~Hu,
\newblock \emph{IEEE Electron Device Letters} \textbf{2002}, \emph{23}, 6 342.

\bibitem{zhou2022experimental}
L.~Zhou, J.~Xiang, X.~Wang, Y.~Zhang, W.~Wang, S.~Feng,
\newblock \emph{Applied Physics A} \textbf{2022}, \emph{128}, 8 730.

\bibitem{monch2007electric}
W.~M{\"o}nch,
\newblock \emph{Applied Physics Letters} \textbf{2007}, \emph{91}, 4.

\bibitem{lu2010bilayer}
C.-H. Lu, G.~M. Wong, R.~Birringer, R.~Dauskardt, M.~D. Deal, B.~M. Clemens, Y.~Nishi,
\newblock \emph{Journal of Applied Physics} \textbf{2010}, \emph{107}, 6.

\bibitem{maurand2016cmos}
R.~Maurand, X.~Jehl, D.~Kotekar-Patil, A.~Corna, H.~Bohuslavskyi, R.~Lavi{\'e}ville, L.~Hutin, S.~Barraud, M.~Vinet, M.~Sanquer, et~al.,
\newblock \emph{Nature Communications} \textbf{2016}, \emph{7}, 1 13575.

\bibitem{kim2023high}
S.~J. Kim, J.~W. Jeong, S.~Y. Park, C.~W. Chung,
\newblock \emph{Materials Science and Engineering: B} \textbf{2023}, \emph{293} 116494.

\bibitem{niebojewski2022specificities}
H.~Niebojewski, B.~Bertrand, E.~Nowak, T.~B{\'e}d{\'e}carrats, B.~C. Paz, L.~Contamin, P.-A. Mortemousque, V.~Labracherie, L.~Brevard, H.~Sahin, et~al.,
\newblock In \emph{2022 IEEE Symposium on VLSI Technology and Circuits (VLSI Technology and Circuits)}. IEEE, \textbf{2022} 415--416.

\bibitem{saraiva2022materials}
A.~Saraiva, W.~H. Lim, C.~H. Yang, C.~C. Escott, A.~Laucht, A.~S. Dzurak,
\newblock \emph{Advanced Functional Materials} \textbf{2022}, \emph{32}, 3 2105488.

\bibitem{elsayed2024low}
A.~Elsayed, M.~Shehata, C.~Godfrin, S.~Kubicek, S.~Massar, Y.~Canvel, J.~Jussot, G.~Simion, M.~Mongillo, D.~Wan, et~al.,
\newblock \emph{npj Quantum Inf} \textbf{2024}, \emph{10}, 70.

\bibitem{thomas2008electron}
J.~M. Thomas, E.~T. Simpson, T.~Kasama, R.~E. Dunin-Borkowski,
\newblock \emph{Accounts of Chemical Research} \textbf{2008}, \emph{41}, 5 665.

\bibitem{boureau_off-axis_2018}
V.~Boureau, R.~McLeod, B.~Mayall, D.~Cooper,
\newblock \emph{Ultramicroscopy} \textbf{2018}, \emph{193} 52.

\bibitem{vansteenkiste2014design}
A.~Vansteenkiste, J.~Leliaert, M.~Dvornik, M.~Helsen, F.~Garcia-Sanchez, B.~Van~Waeyenberge,
\newblock \emph{AIP Advances} \textbf{2014}, \emph{4}, 10.

\bibitem{michaelson1977work}
H.~B. Michaelson,
\newblock \emph{Journal of Applied Physics} \textbf{1977}, \emph{48}, 11 4729.

\end{thebibliography}


\newpage

\newpage
\newpage

\newpage

\begin{table}
\centering
 \caption{Results from C(f)-V characterization of different gate/oxide combinations before and after thermal annealing in forming gas. A comparison is given between non-magnetic (Ti-Pd) and magnetic (Cr-Co) gates.}
  \begin{tabular}[htbp]{@{}lllllll@{}}
    \hline
    Oxide/Metal & Annealing T (°C) & t$_{ox}$(nm) & V$_{FB}$(V) & $\phi_s$(eV) & $\phi_{m, eff}$ (eV) & N$_{eff}$(cm$^{-3}$) \\
    \hline
    Thermal SiO$_2$ - Ti/Pd  & -  & 5/10 & -1.1 & 4.91 & 3.8 & 3$\cdot10^{10}$\\
    Al$_2$O$_3$ - Ti/Pd  & -  & 10/15 & 0.73 & 4.92 & 5.4 & 3$\cdot10^{11}$\\
    Al$_2$O$_3$ - Ti/Pd  & 300  & 10/15 & 0.74 & 4.93 & 5.7 & 1.7$\cdot10^{11}$\\
    Thermal SiO$_2$ - Cr/Co & -  & 5/10 & -0.63 & 4.92 & 4.3 & 8$\cdot10^{10}$\\
    Al$_2$O$_3$ - Cr/Co  & -  & 10/15 & 0.48 & 4.91 & 4.8 & 5.9$\cdot10^{11}$\\
    Al$_2$O$_3$ - Cr/Co  & 300  & 10/15 & 0.6 & 4.91 & 5.5 & 3$\cdot10^{10}$\\
    \hline
    \end{tabular}\\
    Vacuum workfunctions \cite{michaelson1977work}: $\phi_{Ti} = 4.33 \; \mathrm{eV}$, $\phi_{Pd} = 5.12\;  \mathrm{eV}$, $\phi_{Cr} = 4.5\;  \mathrm{eV}$, $\phi_{Co} = 5.0 \; \mathrm{eV}$  
\end{table}

\end{document}